\documentclass[a4paper,12pt,twoside]{article}

\usepackage[dvips]{graphicx}
\usepackage{geometry}
\usepackage[centertags]{amsmath}
\usepackage{amsfonts}
\usepackage{amssymb}
\usepackage{latexsym}
\usepackage{mathrsfs}
\usepackage{fancyhdr}
\usepackage[sort&compress,comma,square]{natbib}

\RequirePackage[pdfstartview=FitH]{hyperref}

\hypersetup{%
  pdftitle={Action Principle and Algebraic Approach to
  Gauge Transformations in Gauge Theories},%
  pdfsubject={Gauge Field Theories},%
  pdfauthor={Edouard Berge Manoukian and Suppiya Siranan},%
  pdfkeywords={Schwinger (Quantum) Action Principle,
  Gauge Transformations, vacuum-to-vacuum transition amplitude},%
}

\newcommand{\email}[1]{\href{mailto:#1}{\texttt{#1}}}

\newcommand{\braket}[2]{\ensuremath{\left\langle{#1}\!%
\mathrel{\left|{\vphantom{{#1} {#2}}}\right.%
\kern-\nulldelimiterspace}\!{#2}\right\rangle}}

\newcommand{\ketbra}[2]{\ensuremath{\left|{#1}\!%
\mathrel{\left\rangle\vphantom{{#1} {#2}}\right\langle%
\kern-\nulldelimiterspace}\!{#2}\right|}}

\newcommand{\BK}[3]{\ensuremath{\left\langle{#1}\!%
\mathrel{\left|\vphantom{{#1}}{#2}\vphantom{{#3}}\right|%
\kern-\nulldelimiterspace}\!{#3}\right\rangle}}

\newcommand{\Vacuum}{\ensuremath{\braket{0_{+}}{0_{-}}}}

\newcommand{\uI}{\ensuremath{\mathrm{i}}}
\newcommand{\uE}{\ensuremath{\mathrm{e}}}
\newcommand{\uD}{\ensuremath{\mathrm{d}}}

\renewcommand{\Vec}[1]{\ensuremath{\boldsymbol{\vec{\mathrm{#1}}}}}

\def\Beginboxit{\par\vbox\bgroup\hrule\hbox\bgroup%
                \vrule \kern1.2pt \vbox\bgroup\kern1.2pt}
\def\Endboxit{\kern1.2pt\egroup\kern1.2pt\vrule\egroup%
              \hrule\egroup}{}
\newenvironment{boxit}{\Beginboxit}{\Endboxit}
\newenvironment{boxit*}{\Beginboxit\hbox to\hsize{}}{\Endboxit}
\newcommand{\ArXivSite}{http://arxiv.org/abs/}
\newcommand{\MyArXivNo}{0706.1631}
\newcommand{\ArXivNo}[1]{\href{\ArXivSite#1}{#1}}

\newcommand{\XXXSize}{{\fontsize{12}{12}\selectfont\fbox{%
\textbf{arXiv:\ArXivNo{\MyArXivNo}v2 [hep-th]}}}}
\newcommand{\XXXTitle}{\hfill\XXXSize\newline\vskip 0.4cm}

\makeatletter
\let\@afterindentfalse\@afterindenttrue
\@afterindenttrue \makeatother
\begin{document}
\pagestyle{fancy}

\allowdisplaybreaks

\sloppy

\title{\XXXTitle\textbf{Action Principle and Algebraic Approach to Gauge Transformations
in Gauge Theories}\footnote{Published in \textit{International Journal
of Theoretical Physics}, Vol.~\textbf{44}, No.~1, pp.~53--62
(2005). \
[doi:\href{http://dx.doi.org/10.1007/s10773-005-1436-z}%
{10.1007/s10773-005-1436-z}] %
}}

\author{\textsc{Edouard~B.~Manoukian}\footnote{E-mail:~\email{edouard@sut.ac.th}} \ and
\ \textsc{Suppiya~Siranan} \\
{\href{http://physics3.sut.ac.th/}{School of Physics}, \ %
\href{http://www.sut.ac.th/science/}{Institute of Science}} \\
{\href{http://www.sut.ac.th/}{Suranaree University of Technology}} \\
{Nakhon~Ratchasima, 30000, Thailand}}

\date{}

\maketitle

\begin{boxit}
\begin{abstract}
The action principle is used to derive, by an entirely
\emph{algebraic} approach, gauge transformations of the full
vacuum-to-vacuum transition amplitude (generating functional) from
the Coulomb gauge to arbitrary covariant gauges and in turn to the
celebrated Fock--Schwinger (FS) gauge for the abelian (QED) gauge
theory without recourse to path integrals or to commutation rules
and without making use of delta functionals.  The interest in the
FS gauge, in particular, is that it leads to Faddeev--Popov
ghosts-free non-abelian gauge theories.  This method is expected
to be applicable to non-abelian gauge theories including
supersymmetric ones. \\

\noindent\textbf{PACS Numbers:} \texttt{11.15.-q, 11.15.Bt,
11.15.Tk, 12.20.-m}
\end{abstract}
\end{boxit}

\section{Introduction}

About two decades ago, we have seen
\cite{Manoukian_1986,Manoukian_1987} that the very elegant action
principle \cite{Schwinger_1951a,Schwinger_1951b,Schwinger_1953a,%
Schwinger_1953b,Schwinger_1954} may be
used to quantize gauge theories in constructing the
vacuum-to-vacuum transition amplitude and the Faddeev--Popov
factor \cite{Faddeev_1967}, encountered in non-abelian gauge
theories, was obtained \emph{directly} from the action principle
without much effort. No appeal was made to path integrals, no
commutation rules were used, and there was not even the need to go
into the well known complicated structure of the Hamiltonian
\cite{Fradkin_1970} in non-abelian gauge theories. Of course path
integrals are extremely useful in many respects and may be
formally derived from the action principle
cf.~\cite{Symanzik_1954,Lam_1965,Manoukian_1985}.   We have worked
in the Coulomb gauge, where the physical components are clear at
the outset, to derive the expression for the vacuum-to-vacuum
transition amplitude (generating functional) including the
Faddeev--Popov factor in non-abelian gauge theories.   It is
interesting to note also that the Coulomb gauge naturally arises
\cite{Faddeev_1988,Ogawa_1996}, see also \cite{Joglekar_2002}, in
gauge field theories as constrained dynamics
cf.~\cite{Henneaux_1992,Garcia_1996,Su_2001}. To make transitions
of the generating functional to arbitrary covariant gauges, we
have made use \cite{Manoukian_1986,Manoukian_1987}, in the
process, of so-called $\delta$~functionals \cite{Schwinger_1965}.
The $\delta$~functionals, however, are defined as infinite
dimensional continual integrals corresponding to the different
points of spacetime and hence the gauge transformations were
carried out in the spirit of path integrals.

The purpose of the present investigation is, in particular, to
remedy the above situation involved with delta functionals, and we
here derive the gauge transformations, providing explicit
expressions, for the full vacuum-to-vacuum transition amplitude to
the generating functionals of arbitrary covariant gauges and, in
turn, to the celebrated Fock--Schwinger (FS) gauge
$x^{\mu}A_{\mu}=0$ \cite{Fock_1937,Schwinger_1951c}, as well as
the axial gauge $n^{\mu}A_{\mu}=0$ for a fixed vector $n^{\mu}$,
for the abelian (QED) gauge theory by an entirely \emph{algebraic}
approach dealing only with commuting (or anti-commuting) external
sources.   The interest in the FS gauge, in gauge theories, in
general, is that it leads to Faddeev--Popov ghost-free theories,
cf.~\cite{Kummer_1986}, the gauge field may be expressed quite
simply in terms of the field strength
\cite{Kummer_1986,Durand_1982} and it turns out to be useful in
non-perturbative studies, cf.~\cite{Shifman_1979a,Shifman_1979b}.
Needless to say, the complete expressions of such generating
functionals allow one to obtain gauge transformations of
\emph{all} the Green functions in a theory simply by functional
differentiations with respect to the external sources coupled to
the quantum fields in question and avoids the rather tedious
treatment, but provides information on, the gauge transformation
of diagram by diagram \cite{Handy_1979,Feng_1996} occurring in a
theory.   A key point, whose importance cannot be overemphasized,
in our analysis \cite{Manoukian_1986,Manoukian_1987} is that, a
priori, \emph{no} restrictions are set on the external source(s)
$J^{\mu}$ coupled to the gauge field(s), such as a
$\partial_{\mu}J^{\mu}=0$%
---restriction, so that \emph{variations of the components of
$J^{\mu}$ may be carried out independently}, until the entire
analysis is completed.   The present method is expected to be
applicable to non-abelian gauge theories including supersymmetric
ones and the latter will be attempted in a forthcoming report.
Some classic references which have set the stage of the
investigation of the gauge problem in field theory are given in
\cite{Landau_1954,Landau_1955,Johnson_1959,Zumino_1960,%
Bialynicki-Birula_1968,Mills_1971,Slavnov_1972,Taylor_1971,%
Abers_1973,Wess_1974,Salam_1974,Becchi_1975,Utiyama_1977}.
For more recent studies which are, however, more involved with
field operator techniques and their gauge transformations may be
found in \cite{Sardanashvily_1984,Kobe_1985,Oh_1987,Sugano_1990,%
Gastmans_1996,Pons_1997,Gastmans_1998,Banerjee_2000}.

\section{Gauge Transformations}

The Lagrangian density under consideration is given by a well
known expression \cite{Manoukian_1986,Manoukian_1987}
\begin{align}
  \mathscr{L} =& -\frac{1}{4}F_{\mu\nu}F^{\mu\nu}+\frac{1}{2}\left[\left(
  \frac{\partial_{\mu}\overline{\psi}}{\uI}\right)\gamma^{\mu}\psi
  -\overline{\psi}\gamma^{\mu}\frac{\partial_{\mu}\psi}{\uI}\right]
  -m_{0}\overline{\psi}\psi
  \nonumber \\[0.5\baselineskip]
  &\quad{} +e_{0}\overline{\psi}\gamma_{\mu}\psi{}A^{\mu}
  +\overline{\eta}\psi+\overline{\psi}\eta+A_{\mu}J^{\mu}
  \label{Eqn01}
\end{align}
where $\overline{\eta}$, $\eta$, $J^{\mu}$ are external sources,
and no restriction is set on $J^{\mu}$ (such as
$\partial_{\mu}J^{\mu}=0$) in order to carry out functional
differentiations with respect to all of its components
\emph{independently}.

Our starting point is the vacuum-to-vacuum transition amplitude in
the Coulomb gauge given by \cite{Manoukian_1986,Manoukian_1987}
\begin{align}
  \Vacuum &= \exp\left[\uI\int\!\!\mathscr{L}'_{\mathrm{I}}\right]
  \Vacuum_{0} \equiv{} Z_{\mathrm{C}}\big[\eta,\overline{\eta},J\big]
  \label{Eqn02} \\[0.5\baselineskip]
  \int\!\!\mathscr{L}'_{\mathrm{I}}(\eta,\overline{\eta},J) &=
  \int\!(\uD{}x)\left(e_{0}\frac{\delta}{\uI\delta\eta(x)}\gamma^{\mu}
  \frac{\delta}{\uI\delta\overline{\eta}(x)}
  \frac{\delta}{\uI\delta{}J^{\mu}(x)}\right)
  \label{Eqn03}
\end{align}
where
\begin{align}
  \Vacuum_{0} =& \exp\left[\uI\int\!(\uD{}x)(\uD{}x')\:\overline{\eta}(x)
  S_{+}(x-x')\eta(x')\right]
  \nonumber \\[0.5\baselineskip]
  &\quad{} \times\exp\left[\frac{\uI}{2}\int\!(\uD{}x)(\uD{}x')\:
  J^{\mu}(x)D^{\mathrm{C}}_{\mu\nu}(x,x')J^{\nu}(x')\right]
  \label{Eqn04}
\end{align}
with $S_{+}(x-x')$ denoting the free electron propagator, and, in
the momentum description, ($k,m=1,2,3$),
\begin{align}
  D^{\mathrm{C}}_{km}(q) &= \left(\delta_{km}-\frac{q_{k}q_{m}}{\Vec{q}^{\,2}}\right)
  \frac{1}{q^{2}-\uI\epsilon}
  \label{Eqn05} \\[0.5\baselineskip]
  D^{\mathrm{C}}_{0k}(q) &= 0 = D^{\mathrm{C}}_{k0}(q)
  \label{Eqn06} \\[0.5\baselineskip]
  D^{\mathrm{C}}_{00}(q) &= -\frac{1}{\Vec{q}^{\,2}}.
  \label{Eqn07}
\end{align}

We introduce the generating functional
\begin{align}
  Z\big[\rho,\overline{\rho},K;G\big] =& \exp\left[\uI\int\!\!
  \mathscr{L}'_{\mathrm{I}}(\rho,\overline{\rho},K)\right]
  \nonumber \\[0.5\baselineskip]
  &\quad{} \times\exp\left[\uI\int\!(\uD{}x)(\uD{}x')\:
  \overline{\rho}(x)S_{+}(x-x')\rho(x')\right]
  \nonumber \\[0.5\baselineskip]
  &\quad{} \times\exp\left[\frac{\uI}{2}\int\!(\uD{}x)(\uD{}x')\:
  K_{\mu}(x)D_{G}^{\mu\nu}(x,x')K_{\nu}(x')\right]
  \label{Eqn08}
\end{align}
where in the momentum description
\begin{equation}\label{Eqn09}
  D_{G}^{\mu\nu}(q) = \left(g^{\mu\nu}-\frac{q^{\mu}q^{\nu}}{q^{2}}\right)
  \frac{1}{q^{2}-\uI\epsilon}+q^{\mu}q^{\nu}G(q^{2})
\end{equation}
and $G(q^{2})$ is arbitrary.

We show that
\begin{equation}\label{Eqn10}
  Z_{\mathrm{C}}\big[\eta,\overline{\eta},J\big] = \uE^{\uI{}W'}
  Z\big[\rho,\overline{\rho},K;G\big]\bigg|_{\rho=0,\overline{\rho}=0,K=0}
\end{equation}
where
\begin{align}
  W' &= \int\!(\uD{}x)\:\overline{\eta}(x)\exp\left[-\uI{}e_{0}a^{\mu}
  \frac{\delta}{\uI\delta{}K^{\mu}(x)}\right]
  \frac{\delta}{\uI\delta\overline{\rho}(x)}
  \nonumber \\[0.5\baselineskip]
  &\quad{} +\int\!(\uD{}x)\:\frac{\delta}{\uI\delta\rho(x)}
  \exp\left[\uI{}e_{0}a^{\mu}\frac{\delta}{\uI\delta{}K^{\mu}(x)}\right]\eta(x)
  \nonumber \\[0.5\baselineskip]
  &\quad{} +\int\!(\uD{}x)\:\Big(\big(g^{\mu\sigma}-a^{\mu}\partial^{\sigma}\big)
  J_{\sigma}(x)\Big)\frac{\delta}{\uI\delta{}K^{\mu}(x)}
  \label{Eqn11}
\end{align}
and
\begin{equation}\label{Eqn12}
  a^{\mu} = \left(0\;,\;\frac{\Vec{\nabla}}{\nabla^{2}}\right)
  = g^{\mu{}k}\frac{\partial^{k}}{\nabla^{2}}
\end{equation}
relating the Coulomb gauge to arbitrary covariant gauges.

To establish (\ref{Eqn10}), we start from its right-hand side. We
note, in a matrix notation, that
\begin{align}
  &\uE^{\uI{}W'}\exp\big[\uI\overline{\rho}S_{+}\rho\big]\exp\left[
  \frac{\uI}{2}K_{\mu}D_{G}^{\mu\nu}K_{\nu}\right]
  \nonumber \\[0.5\baselineskip]
  &\quad{} =\exp\left[\uI\left(\overline{\rho}+\overline{\eta}\exp\left[
  -\uI{}e_{0}a^{\mu}\frac{\delta}{\uI\delta{}K^{\mu}}\right]\right)S_{+}
  \left(\rho+\exp\left[\uI{}e_{0}a^{\mu}\frac{\delta}{\uI\delta{}K^{\mu}}
  \right]\eta\right)\right]
  \nonumber \\[0.5\baselineskip]
  &\quad\qquad{} \times\exp\left[\frac{\uI}{2}\Big(K_{\mu}+\big(g_{\mu\sigma}
  -a_{\mu}\partial_{\sigma}\big)J^{\sigma}\Big)D_{G}^{\mu\nu}
  \Big(K_{\nu}+\big(g_{\nu\lambda}-a_{\nu}\partial_{\lambda}\big)J^{\lambda}\Big)\right]
  \label{Eqn13}
\end{align}
and since $\mathscr{L}'_{\mathrm{I}}(\rho,\overline{\rho},K)$, is
classical, is invariant under transformations
$\rho(x)\to\rho(x)\exp\big(\uI\Lambda(x)\big)$,
$\overline{\rho}(x)\to\exp\big({-\uI}\Lambda(x)\big)\overline{\rho}(x)$
for an arbitrary numerical function $\Lambda(x)$, and we
eventually set $\rho=0$, $\overline{\rho}=0$, the right-hand side
of (\ref{Eqn10}) becomes
\begin{align}
  &\exp\left[\uI\int\!\!\mathscr{L}'_{\mathrm{I}}(\eta,\overline{\eta},J)\right]
  \nonumber \\[0.5\baselineskip]
  &\quad{} \times\exp\left[\uI\left(\overline{\eta}\exp\left[-\uI{}e_{0}a^{\mu}
  \frac{\delta}{\uI\delta{}K^{\mu}}\right]\right)S_{+}\left(
  \exp\left[\uI{}e_{0}a^{\mu}\frac{\delta}{\uI\delta{}K^{\mu}}\right]
  \eta\right)\right]
  \nonumber \\[0.5\baselineskip]
  &\quad{} \times\exp\left[\frac{\uI}{2}\Big(K_{\mu}+\big(g_{\mu\sigma}
  -a_{\mu}\partial_{\sigma}\big)J^{\sigma}\Big)D_{G}^{\mu\nu}
  \Big(K_{\nu}+\big(g_{\nu\lambda}-a_{\nu}\partial_{\lambda}\big)J^{\lambda}\Big)\right]
  \label{Eqn14}
\end{align}
with $K_{\mu}\to{}0$.   Now we use the identity
\begin{align}
  &\exp\left[\uI{}e_{0}\int\!(\uD{}x)\left(\frac{\delta}{\uI\delta\eta(x)}
  \gamma^{\mu}\frac{\delta}{\uI\delta\overline{\eta}(x)}
  \partial_{\mu}\Lambda(x)\right)\right]\exp\big[\uI\overline{\eta}S_{+}\eta\big]
  \nonumber \\[0.5\baselineskip]
  &\qquad{} =\exp\Big[\uI\left(\overline{\eta}\,\uE^{\uI{}e_{0}\Lambda}\right)S_{+}
  \left(\uE^{{-\uI{}}e_{0}\Lambda}\eta\right)\Big]
  \label{Eqn15}
\end{align}
to rewrite the above expression as
\begin{align}
  &\exp\left[\uI{}e_{0}\int\!(\uD{}x)\left(\frac{\delta}{\uI\delta\eta(x)}
  \gamma_{\mu}\frac{\delta}{\uI\delta\overline{\eta}(x)}
  \big(g^{\mu\sigma}-a^{\mu}\partial^{\sigma}\big)
  \frac{\delta}{\uI\delta{}K^{\sigma}(x)}\right)\right]
  \exp\big[\uI\overline{\eta}S_{+}\eta\big]
  \nonumber \\[0.5\baselineskip]
  &\quad{} \times\exp\left[\frac{\uI}{2}\Big(K_{\mu}+\big(g_{\mu\sigma}
  -a_{\mu}\partial_{\sigma}\big)J^{\sigma}\Big)D_{G}^{\mu\nu}
  \Big(K_{\nu}+\big(g_{\nu\lambda}-a_{\nu}\partial_{\lambda}\big)J^{\lambda}\Big)\right]
  \label{Eqn16}
\end{align}
which for $K_{\mu}\to{}0$ reduces to the left-hand side of
(\ref{Eqn10}) \emph{since}
\begin{equation}\label{Eqn17}
  \big(g_{\mu\sigma}-a_{\mu}\partial_{\sigma}\big)D_{G}^{\mu\nu}
  \big(g_{\nu\lambda}-a_{\nu}\partial_{\lambda}\big) =
  D^{\mathrm{C}}_{\sigma\lambda}.
\end{equation}

Almost an identical analysis as above shows, by noting in the
process,
\begin{equation}\label{Eqn18}
  \big(g_{\mu\sigma}-\widetilde{a}_{\mu}\partial_{\sigma}\big)D_{G}^{\mu\nu}
  \big(g_{\nu\lambda}-\widetilde{a}_{\nu}\partial_{\lambda}\big)
  = \big(D_{0}\big)_{\sigma\lambda} \equiv{} D^{\mathrm{L}}_{\sigma\lambda}
\end{equation}
with
\begin{equation}\label{Eqn19}
  \widetilde{a}_{\mu} = \dfrac{\partial_{\mu}}{\square},\qquad
  \square\equiv\partial_{\mu}\partial^{\mu}
\end{equation}
where the right-hand side of (\ref{Eqn18}) defines the photon
propagator in the Landau gauge, with $G$ in (\ref{Eqn09}) set
equal to zero, that
\begin{equation}\label{Eqn20}
  Z\big[\eta,\overline{\eta},J;G=0\big] = \uE^{\uI\widetilde{W}'}
  Z\big[\rho,\overline{\rho},K;G\big]\bigg|_{\rho=0,\overline{\rho}=0,K=0}
\end{equation}
where $\widetilde{W}'$ is given by the expression defined in
(\ref{Eqn11}) with $a^{\mu}$ in it simply replaced by
$\widetilde{a}^{\mu}$, thus relating the Landau gauge to arbitrary
covariant gauges.

The Fock--Schwinger gauge $x^{\mu}A_{\mu}=0$, allows one to write
\begin{equation}\label{Eqn21}
  A^{0} = \frac{x^{k}A_{k}}{x^{0}}
\end{equation}
which upon substitution in (\ref{Eqn01}), and varying
$\mathscr{L}$ with respect to $A^{k}$ yields
\begin{equation}\label{Eqn22}
  \partial_{\mu}F^{\mu{}k}-\frac{x^{k}}{x^{0}}\partial_{\mu}F^{\mu{}0}
  = -j^{k}+j^{0}\frac{x^{k}}{x^{0}}
\end{equation}
where
\begin{equation}\label{Eqn23}
  j^{\mu} = e_{0}\overline{\psi}\gamma^{\mu}\psi+J^{\mu}.
\end{equation}
We note that (\ref{Eqn22}) holds true with $k$ replaced by $0$ in
it giving $0=0$, i.e., we may rewrite (\ref{Eqn22}) as
\begin{equation}\label{Eqn24}
  \partial_{\mu}F^{\mu\nu}-\frac{x^{\nu}}{x^{0}}\partial_{\mu}F^{\mu{}0}
  = -j^{\nu}+j^{0}\frac{x^{\nu}}{x^{0}} \equiv{} S^{\nu}.
\end{equation}

By taking the derivative $\partial_{\nu}$ of (\ref{Eqn24}), we may
solve for $\left(\partial_{\mu}F^{\mu{}0}\right)/x^{0}$,
\begin{equation}\label{Eqn25}
  -\frac{\partial_{\mu}F^{\mu{}0}}{x^{0}} = \big(\partial\,x\big)^{-1}\partial_{\sigma}
  \left(-j^{\sigma}+j^{0}\frac{x^{\sigma}}{x^{0}}\right)
\end{equation}
which upon substituting in (\ref{Eqn24}) gives
\begin{equation}\label{Eqn26}
  \partial_{\mu}F^{\mu\nu} = -\Big[g^{\nu\sigma}-x^{\nu}
  \big(\partial\,x\big)^{-1}\partial^{\sigma}\Big] j_{\sigma}.
\end{equation}

By taking $\nu=k$, and taking the derivative $\partial_{k}$ of
(\ref{Eqn26}), we may write
\begin{equation}\label{Eqn27}
  -\partial_{0}A^{0} = \frac{1}{\nabla^{2}}
  \Big(\partial_{0}^{2}\,\partial_{k}A^{k}+\partial_{k}S^{k}\Big)
\end{equation}
which when substituted in (\ref{Eqn26}) gives
\begin{equation}\label{Eqn28}
  A^{\nu} = \square^{-1}S^{\nu}+\frac{\partial^{\nu}}{\nabla^{2}}
  \left(\partial_{k}A^{k}-\frac{1}{\square}\partial_{k}S^{k}\right).
\end{equation}
That is, $A^{\nu}$ is of the form
\begin{equation}\label{Eqn29}
  A^{\nu} = \square^{-1}S^{\nu}+\partial^{\nu}a.
\end{equation}
For $\nu=k$, and multiplying (\ref{Eqn29}) by $x^{k}/x^{0}$, we
have from (\ref{Eqn21})
\begin{equation}\label{Eqn30}
  A^{0} = \frac{x^{k}}{x^{0}}\square^{-1}S^{k}+\frac{x^{k}}{x^{0}}\partial^{k}a.
\end{equation}
On the other hand, directly from (\ref{Eqn29}) with $\nu=0$ in it,
\begin{equation}\label{Eqn31}
  A^{0} = \square^{-1}S^{0}+\partial^{0}a
\end{equation}
which upon comparison with (\ref{Eqn30}) leads to
\begin{equation}\label{Eqn32}
  x\,\partial\:a = -x^{\mu}\square^{-1}S_{\mu}.
\end{equation}

From (\ref{Eqn29}), (\ref{Eqn32}) and the definition of $S^{\nu}$
in (\ref{Eqn24}), we obtain
\begin{equation}\label{Eqn33}
  A^{\nu} = -\frac{1}{\square}\left(g^{\nu\mu}-\partial^{\nu}
  \frac{1}{x\,\partial+2}x^{\mu}\right)\left(g_{\mu\sigma}-x_{\mu}
  \frac{1}{\partial\,x}\partial_{\sigma}\right)j^{\sigma}
\end{equation}
where we have noted that $\partial\,x=4+x\,\partial$.   It is
straightforward to check from (\ref{Eqn33}) that
$x_{\nu}A^{\nu}=0$ is indeed satisfied.

To establish the transformation from covariant gauges to the FS
gauge, we have to pull $\square^{-1}$ in (\ref{Eqn33}) between the
two round brackets.   To this end we note that
\begin{equation}\label{Eqn34}
  \square\:x\,\partial = \big(x\,\partial+2\big)\,\square
\end{equation}
and hence
\begin{equation}\label{Eqn35}
  \big(\square\:x\,\partial\big)^{-1} = \big(x\,\partial\big)^{-1}\square^{-1}
  = \square^{-1}\big(x\,\partial+2\big)^{-1}
\end{equation}
i.e.,
\begin{equation}\label{Eqn36}
  \frac{1}{\square}\:\frac{1}{x\,\partial+2} = \frac{1}{x\,\partial}\:\frac{1}{\square}.
\end{equation}
We may also use the identity
\begin{equation}\label{Eqn37}
  \frac{1}{\square}x^{\mu} = x^{\mu}\frac{1}{\square}-2\frac{\partial^{\mu}}{\square}
\end{equation}
and since $\partial^{\mu}$ when applied to the second factor in
(\ref{Eqn33}) gives
\begin{equation}\label{Eqn38}
  \partial^{\mu}\left(g_{\mu\sigma}-x_{\mu}\frac{1}{\partial\,x}\partial_{\sigma}\right)
  = 0.
\end{equation}

We obtain from (\ref{Eqn36})--(\ref{Eqn38}), (\ref{Eqn33})
\begin{equation}\label{Eqn39}
  A^{\nu} = \left(g^{\nu\mu}-\partial^{\nu}\frac{1}{x\,\partial}x^{\mu}\right)
  \frac{1}{\big({-\square}\big)}
  \left(g_{\mu\sigma}-x_{\mu}\frac{1}{\partial\,x}\partial_{\sigma}\right)j^{\sigma}.
\end{equation}

Now we invoke the transversality property in (\ref{Eqn38}) to
\emph{rewrite} (\ref{Eqn39}) as
\begin{equation}\label{Eqn40}
  A^{\nu} = \left(g^{\nu\mu}-\partial^{\nu}\frac{1}{x\,\partial}x^{\mu}\right)
  \frac{1}{\big({-\square}\big)}
  \Big[g_{\mu\rho}-H(\square)\partial_{\mu}\partial_{\rho}\Big]
  \left(g^{\rho\sigma}-x^{\rho}\frac{1}{\partial\,x}\partial^{\sigma}\right)j_{\sigma}
\end{equation}
where $H(\square)$ is \emph{arbitrary} on account of
(\ref{Eqn38}).

It remains to set
\begin{equation}\label{Eqn41}
  g^{\rho\sigma}-x^{\rho}\frac{1}{\partial\,x}\partial^{\sigma} =
  O^{\rho\sigma}
\end{equation}
and note that for the factor multiplying $j_{\sigma}$ on the
right-hand side of (\ref{Eqn40}),
\begin{equation}\label{Eqn42}
  \BK{x}{(\bullet)}{x'} = \int\!\big(\uD{}x''\big)\big(\uD{}x'''\big)
  \BK{x''}{O^{\mu\nu}}{x}\BK{x''}{\big(D_{H}\big)_{\mu\rho}}{x'''}
  \BK{x'''}{O^{\rho\sigma}}{x'}
\end{equation}
where, as shown in the Appendix, we have noted that
\begin{equation}\label{Eqn43}
  \BK{x}{\partial^{\nu}\big(x\,\partial\big)^{-1}x^{\mu}}{x'} =
  \BK{x'}{x^{\mu}\big(\partial\,x\big)^{-1}\partial^{\nu}}{x}
\end{equation}
and we recognize $\BK{x''}{\big(D_{H}\big)_{\mu\rho}}{x'''}$ to
have the very general structure in (\ref{Eqn09}).   Hence we may
write, as in (\ref{Eqn10}),
\begin{equation}\label{Eqn44}
  Z_{\mathrm{FS}}\big[\eta,\overline{\eta},J\big] = \uE^{\uI{}W''}
  Z\big[\rho,\overline{\rho},K;G\big]\bigg|_{\rho=0,\overline{\rho}=0,K=0}
\end{equation}
where $W''$ is given by (\ref{Eqn11}) with $a^{\mu}$ in the latter
replaced by $x^{\mu}\big(\partial\,x\big)^{-1}$.    [For
interpretation of
$x^{\mu}\big(\partial\,x\big)^{-1}\partial^{\nu}$ see the Appendix
and also \cite{Kummer_1986}.]

The axial gauge $n^{\mu}A_{\mu}=0$, with $n^{\nu}$ a fixed vector,
is handled similarly, with $A^{\nu}$ in (\ref{Eqn39}) now replaced
by
\begin{equation}\label{Eqn45}
  A^{\nu} = \left(g^{\nu\mu}-\partial^{\nu}\frac{1}{n\,\partial}n^{\mu}\right)
  \frac{1}{\big({-\square}\big)}
  \left(g_{\mu\sigma}-n_{\mu}\frac{1}{n\,\partial}\partial_{\sigma}\right)j^{\sigma}
\end{equation}
and a similar expression as in (\ref{Eqn44}) holds with $a^{\mu}$
in (\ref{Eqn10}) replaced by $n^{\mu}\big(n\,\partial\big)^{-1}$
in it.

\section{Conclusion}
We have seen that the algebraic method developed in this work
solves the gauge transformation problem relating generating
functionals in different gauges starting from the vacuum-to-vacuum
transition amplitude in the Coulomb gauge.   Needless to say,
their transformation rules give the transformations of \emph{all}
the Green functions encountered in the theory and avoids
unnecessary tedious steps otherwise involved.   The simplicity and
the power of the method is evident and it is expected to be
applicable to non-abelian gauge theories, with
\cite{Manoukian_1986,Manoukian_1987} or without Faddeev--Popov
ghosts, as well as to supersymmetric theories. We have not,
however, touched upon uniqueness problems such as the Gribov
ambiguity \cite{Gribov_1978,Zwanziger_1981}.     This and
extensions to non-abelian cases and supersymmetric theories will
be attempted in a forthcoming report.

\section*{Acknowledgment}
The authors would like to acknowledge with thanks for being
granted \href{http://rgj.trf.or.th/}{the ``Royal Golden Jubilee
Ph.D. Program''} by \href{http://www.trf.or.th/}{the Thailand
Research Fund} (Grant No. PHD/0193/2543) for partly carrying out
this project.

\section*{Appendix}
\renewcommand{\theequation}{A.\arabic{equation}}
\setcounter{equation}{0}

For an explicit derivation of (\ref{Eqn43}), we multiply
$\partial^{\nu}$ by $-\uI$ and write
\begin{equation}\label{EqnA.1}
  \partial^{\nu}\big(x\,\partial\big)^{-1}x^{\mu} = (xp+1)^{-1}p^{\nu}x^{\mu} =
  \sum\limits_{n=0}^{\infty}(-1)^{n}(xp)^{n}p^{\nu}x^{\mu}
\end{equation}
upon moving, in the process, $p^{\nu}$ to the right.   Using the
identity
\begin{equation}\label{EqnA.2}
  \left(x^{\mu}p_{\mu}\right)_{\mathrm{op}} = \int\!(\uD{}x)\frac{(\uD{}p)}{(2\pi)^{4}}
  \ketbra{x}{p} xp \;\uE^{\uI{}xp}
\end{equation}
we note that
\begin{align}
  \big(xp\big)^{n} &= \int\left[\prod\limits_{i=1}^{n}\big(\uD{}x_{i}\big)
  \frac{\big(\uD{}p_{i}\big)}{(2\pi)^{4}}x_{i}p_{i}\right]
  \nonumber \\[0.5\baselineskip]
  &\quad\qquad{} \times\uE^{\uI{}x_{n}(p_{n}-p_{n-1})}
  \uE^{\uI{}x_{n-1}(p_{n-1}-p_{n-2})}\ldots
  \uE^{\uI{}x_{1}p_{1}}\ketbra{x_{1}}{p_{n}}
  \label{EqnA.3}
\end{align}
and hence
\begin{align}
  \BK{x}{\partial^{\nu}\big(x\,\partial\big)^{-1}x^{\mu}}{x'} &=
  \sum\limits_{n=0}^{\infty}(-1)^{n}\!\!
  \int\left[\prod\limits_{i=1}^{n}\big(\uD{}x_{i}\big)
  \frac{\big(\uD{}p_{i}\big)}{(2\pi)^{4}}x_{i}p_{i}\right]p_{n}^{\nu}x'^{\mu}
  \delta(x-x_{1})
  \nonumber \\[0.5\baselineskip]
  &\qquad{} \times\uE^{\uI{}x_{n}(p_{n}-p_{n-1})}\uE^{\uI{}x_{n-1}(p_{n-1}-p_{n-2})}
  \ldots\uE^{\uI{}x_{1}p_{1}}\uE^{-\uI{}p_{n}x}.
  \label{EqnA.4}
\end{align}
This may be rewritten in an equivalent form by making the change
of variables
\begin{equation}\label{EqnA.5}
  x_{1}=y_{n},\ldots,x_{n}=y_{1}\;;\quad{}
  p_{1}=-q_{n},\ldots,p_{n}=-q_{1}
\end{equation}
leading to
\begin{align}
  \BK{x}{\partial^{\nu}\big(x\,\partial\big)^{-1}x^{\mu}}{x'} &=
  -\sum\limits_{n=0}^{\infty}\int\left[\prod\limits_{i=1}^{n}\big(\uD{}y_{i}\big)
  \frac{\big(\uD{}q_{i}\big)}{(2\pi)^{4}}y_{i}q_{i}\right]x'^{\mu}q_{1}^{\nu}
  \delta(y_{n}-x)
  \nonumber \\[0.5\baselineskip]
  &\quad\qquad{} \times\uE^{\uI{}xq_{1}}\uE^{\uI{}y_{1}(q_{2}-q_{1})}
  \uE^{\uI{}y_{2}(q_{3}-q_{2})}\ldots\uE^{-\uI{}y_{n}q_{n}}.
  \label{EqnA.6}
\end{align}

On the other hand,
\begin{align}
  \BK{x}{x^{\mu}\big(\partial\,x\big)^{-1}\partial^{\nu}}{x'} &=
  \BK{x}{x^{\mu}p^{\nu}\big(p\,x-1\big)^{-1}}{x'}
  \nonumber \\[0.5\baselineskip]
  &= -\sum\limits_{n=0}^{\infty}\BK{x}{x^{\mu}p^{\nu}\big(p\,x\big)^{n}}{x'}
  \label{EqnA.7}
\end{align}
and
\begin{equation}\label{EqnA.8}
  \left(p^{\mu}x_{\mu}\right)_{\mathrm{op}} = \int\!(\uD{}x)\frac{(\uD{}p)}{(2\pi)^{4}}
  \ketbra{p}{x} p\,x \;\uE^{-\uI{}px}
\end{equation}
\begin{align}
  \big(p\,x\big)^{n} &= \int\left[\prod\limits_{i=1}^{n}\big(\uD{}x_{i}\big)
  \frac{\big(\uD{}p_{i}\big)}{(2\pi)^{4}}p_{i}x_{i}\right]
  \nonumber \\[0.5\baselineskip]
  &\quad\qquad{} \times\uE^{\uI{}x_{1}(p_{2}-p_{1})}\ldots
  \uE^{\uI{}x_{n-1}(p_{n}-p_{n-1})}
  \uE^{-\uI{}x_{n}p_{n}}\ketbra{p_{1}}{x_{n}}
  \label{EqnA.9}
\end{align}
leading to
\begin{align}
  \BK{x}{x^{\mu}\big(\partial\,x\big)^{-1}\partial^{\nu}}{x'} &=
  -\sum\limits_{n=0}^{\infty}\int\left[\prod\limits_{i=1}^{n}\big(\uD{}x_{i}\big)
  \frac{\big(\uD{}p_{i}\big)}{(2\pi)^{4}}p_{i}x_{i}\right]x^{\mu}p_{1}^{\nu}
  \delta(x_{n}-x')
  \nonumber \\[0.5\baselineskip]
  &\quad\qquad{} \times\uE^{\uI{}xp_{1}}\uE^{\uI{}x_{1}(p_{2}-p_{1})}\ldots
  \uE^{\uI{}x_{n-1}(p_{n}-p_{n-1})}\uE^{-\uI{}x_{n}p_{n}}
  \label{EqnA.10}
\end{align}
which upon comparison with (\ref{EqnA.6}) establishes
(\ref{Eqn43}).

\end{document}